\documentclass[preprint,12pt]{elsart}

\usepackage{graphicx}
\usepackage{amssymb}
\usepackage{amsmath}
\usepackage{subfigure}

\usepackage{tikz}
\usetikzlibrary{backgrounds}

\usepackage[
    pdftex,
    pdftitle={Removing the Barrier to Scalability in Parallel FMM},
    pdfauthor={Matthew G.~Knepley},
    pdfpagemode={UseOutlines},
    bookmarks, bookmarksopen,bookmarksnumbered={True},
    colorlinks, linkcolor={blue},citecolor={blue},urlcolor={blue}
    ]{hyperref}

\journal{Parallel Computing}

\newcommand{\uKern}{\mathbb{K}}

\begin{document}

\begin{frontmatter}


\title{Removing the Barrier to Scalability in Parallel FMM}


\author[mgk]{Matthew G.~Knepley}
\address[mgk]{Computation Institute, University of Chicago}

\begin{abstract}
    The Fast Multipole Method (FMM) is well known to possess a bottleneck arising from decreasing workload on higher
levels of the FMM tree [Greengard and Gropp, Comp. Math. Appl., 20(7), 1990]. We show that this potential bottleneck can
be eliminated by overlapping multipole and local expansion computations with direct kernel evaluations on the finest
level grid.
\end{abstract}

\end{frontmatter}

\section{Introduction}\label{sec:intro}

    In~\cite{GreengardGropp1990}, Greengard and Gropp give the seminal complexity analysis for the parallel Fast
Multipole Method (FMM)~\cite{GreengardRokhlin1987}. A notable finding is that the algorithm contains a bottleneck which
scales as $\log P$, where $P$ is the number of processors. This kind of bottleneck is found in many other hierarchical
algorithms, such as multigrid~\cite{Brandt77}, and results directly from a lack of concurrency. Work
on a given grid level depends upon results from a finer grid. Although some freedom can be
exploited~\cite{StroutCarterFerranteFreemanKreaseck02}, grid levels are typically executed sequentially, concurrency arising only from
operations on a given level. In their analysis, Greengard and Gropp also make this tacit assumption.

\begin{figure}\label{fig:fmmStages}
\tikzstyle{bg1}       = [scale=0.75,draw=black,fill=gray!20!white,minimum width=4cm,minimum height=8cm]
\tikzstyle{bg2}       = [scale=0.75,draw=black,fill=gray!80!white,minimum width=4cm,minimum height=8cm]
\tikzstyle{operation} = [scale=0.75,circle,draw=black,fill=white,inner sep=0pt,minimum size=1cm]
\tikzstyle{sink}      = [scale=0.75,circle,draw=black,fill=black,inner sep=0pt,minimum size=1cm]
\tikzstyle{dep}       = [<-,shorten <=1pt,ultra thick]
\begin{tikzpicture}[scale=0.75]
  \node at (2,4)  [bg1,label=above:{\small Setup}] {};
  \node at (6,4)  [bg2,label=above:{\small Up Sweep}] {};
  \node at (10,4) [bg1,label=above:{\small Down Sweep}] {};
  \node at (14,4) [bg2,label=above:{\small Evaluation}] {};
  \node (start)         at (-1,4) [sink] {};
  \node (createTree)    at (1,4) [operation] {1}
    edge [dep] (start);
  \node (addParticles)  at (3,6) [operation] {2}
    edge [dep] (createTree);
  \node (calcNeighbors) at (3,2) [operation] {3}
    edge [dep] (createTree);
  \node (P2M)           at (5,6) [operation] {4}
    edge [dep] (addParticles);
  \node (M2M)           at (7,6) [operation] {5}
    edge [dep] (P2M);
  \node (M2L)           at (9,6) [operation] {6}
    edge [dep] (calcNeighbors)
    edge [dep] (M2M);
  \node (L2L)           at (11,6) [operation] {7}
    edge [dep] (M2L);
  \node (L2P)           at (13,6) [operation] {8}
    edge [dep] (L2L);
  \node (calcDirect)    at (13,2) [operation] {9}
    edge [dep] (addParticles)
    edge [dep] (calcNeighbors);
  \node (calcTotal)     at (15,4) [operation] {10}
    edge [dep] (L2P)
    edge [dep] (calcDirect);
  \node (end)           at (17,4) [sink] {}
    edge [dep] (calcTotal);
\end{tikzpicture}

\caption{Directed Acyclic Graph representing the computational stages of the FMM algorithm. Notice that no dependencies
exist between \{3\} and \{2,4,5\}, and \{6,7,8\} and \{9\}. This means that these task groups may be executed
simultaneously. The numbered tasks correspond to: 1. Tree construction 2. Particle binning 3. Interaction list
calculation 4. Particle to multipole calculation 5. Multipole to multipole translation 6. Multipole to local
transformation 7. Local to local translation 8. Local to particle evaluation 9. Near domain particle evaluation
10. Summing multipole and near field contributions.}
\end{figure}
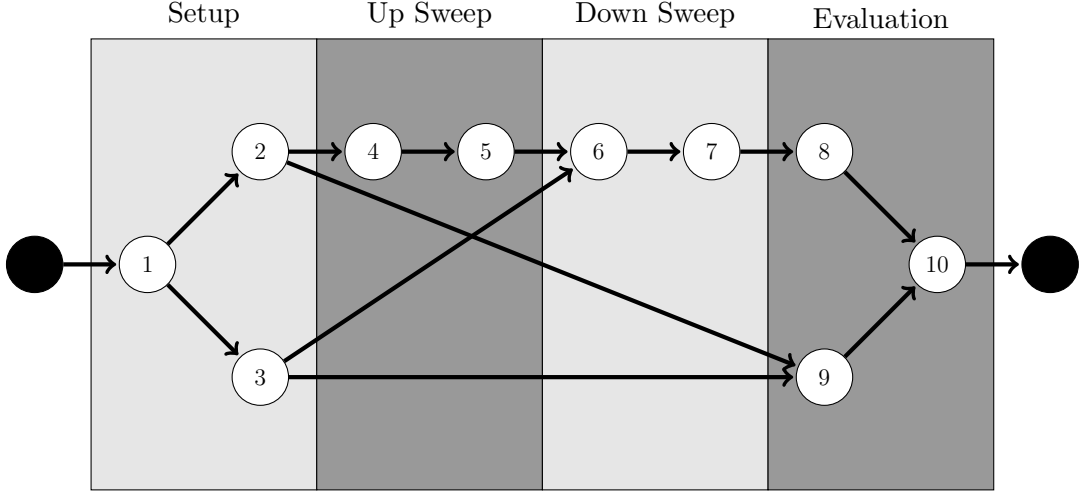

    However, even allowing for serialization of grid levels, we are left with potential concurrency in FMM that is not
accounted for in the complexity model. All local interaction calculations are independent of the multipole calculations,
and also are localized to a given cell and its neighbors. The dependency graph for the computational stages in FMM is
shown in Fig.~\ref{fig:fmmStages}. Notice that no dependency exists between the multipole calculation, occuring in stages
\{4, 5, 6, 7\}, and the direct near-field calculation in stage \{9\}.

If all neighbor data is made available at the start of the computation, these local calculations could be used to offset
the dearth of concurrency at coarser grid levels. In fact, we will show in Section~\ref{sec:concurrency} that for nearly
every architecture in use today, local interaction calculations are sufficient to completely cover the bottleneck in the
case of our test case from vortex fluid dynamics.

\section{Test problem}\label{sec:problem}

    In order to provide concrete values for all coefficients in the complexity estimates, we will focus on a particular
test problem, the Euler equations of fluid dynamics in two dimensions. We will write the equation in the
velocity-vorticity formulation, where vorticity is defined as the curl of the velocity, $\omega = \nabla\times
u$~\cite{Pozrikidis07}. The equations are discretized using the vortex particle method over a set of moving nodes
located at $x_i$ as follows:
\begin{equation}\label{eq:omega}
  \omega(x, t) \approx \omega_{\sigma}(x, t) = \sum_i^N \gamma_i \zeta_{\sigma}(x, x_i(t))
\end{equation}
where a common choice for the basis function $\zeta_{\sigma}(x,y)$ is a normalized Gaussian,
\begin{equation}\label{eq:zeta}
  \zeta_{\sigma}(x,y) = \frac{1}{2\pi\sigma^2} \exp{
  \left( \frac{-|x-y|^2}{2\sigma^2} \right) }.
\end{equation}
This formulation was studied previously in~\cite{CruzBarba2009} for accuracy, and a high performance implementation was
described in~\cite{CruzBarbaKnepley2009}.

    The discretized vorticity field in~\eqref{eq:omega} is used in conjunction with the vorticity transport equation.
For ideal flows in two dimensions, the vorticity is a preserved quantity over particle trajectories,
\begin{equation}\label{eq:2dEuler}
  \frac{\partial \omega}{\partial t} + u\cdot \nabla \omega
  = \frac{ {\rm D}\omega}{ {\rm D} t} = 0,
\end{equation}
and the moving nodes translate with their local velocity, carrying their vorticity value to automatically satisfy the
transport equation. We then calculate velocity from the discretized vorticity using the Biot-Savart law, 
\begin{equation} \label{eq:biotsav}
  u(x,t) = \int (\nabla\times\mathbb{G})(x-x^{\prime}) \omega(x^{\prime},t)dx^{\prime}\nonumber
   = \int\uKern(x-x^{\prime})\omega(x^{\prime},t) dx^{\prime} = (\uKern\ast\omega)(x,t)
\end{equation}
where $\uKern = \nabla\times\mathbb{G}$ is the Biot-Savart kernel obtained from $\mathbb{G}$, the Green's function for
the Poisson equation, and $\ast$ represents convolution. In our 2D example, the Biot-Savart law is written explicitly
as,
\begin{equation} \label{eq:biotsav2d}
  u(x,t) = \frac{-1}{2\pi} \int \frac{\left(x -x^{\prime}\right)\times\omega(x^{\prime},t)\mathbf{\hat {k}}}
    {\left|x-x^{\prime}\right|^{2}} dx^{\prime}.
\end{equation}
where $x = (x_1, x_2)$. However, in two dimensions, we can also conveniently represent the vector $x$ as a complex
number $z = u + i v$, for which we will also use the notation $(u, v)$.

    When the vorticity is discretized using a radial basis function expansion, one can always find the analytic integral
for the Biot-Savart velocity, resulting in an expression for the velocity at each node which is a sum over all
particles. Using the Gaussian basis function~\eqref{eq:zeta},
\begin{equation}\label{eq:regB-S2DGau}
  \uKern_{\sigma}(z) = \frac{1}{2 \pi |z|^{2}}(-v, u) \left(1-\exp\left(-\frac{|z|^{2}}{2\sigma^{2}}\right)\right).
\end{equation}
where $|z|^2=u^2 + v^2$, and the velocity is given by,
\begin{equation}\label{eq:discBiotS2D}
  u_{\sigma}(z,t) = \sum_{j=1}^{N} \gamma_{j}\;\uKern_{\sigma}(z-z_{j}(t)).
\end{equation}
Thus, the calculation of the velocity of $N$ vortex particles is an $N$-body problem, where the kernel $\uKern(z)$
decays quadratically with distance, which makes it a candidate for acceleration using the FMM.  Also note that as $|z|$
becomes large,  the kernel $\uKern(z)$ approaches $1/|z|^2$.  We take advantage of this fact to use the multipole
expansions of the $1/|z|^2$ kernel as an approximation, while the near-field direct interactions are obtained with the
exact kernel $\uKern$.  It has been demonstrated that using the expansions for $1/|z|^2$ does not impair the accuracy
as long as the local interaction boxes are not too small~\cite{CruzBarba2009}.

\section{Complexity Estimates}\label{sec:complexity}

    Greengard and Gropp begin by assuming a constant number $B$ of particles per cell on the finest tree level $L$. In
dimension $D$, as a function of $N = 2^{D L} B$, the total number of particles, and $P$, the number of processes, the
runtime $T$ is given by
\begin{equation}
  \label{eq:complexity}
  T = a \frac{N}{P} + b \log_{2^D} P + c \frac{N}{B P} + d \frac{N B}{P} + e(N, P)
\end{equation}
where $a$--$d$ are constants explained below, and $e$ subsumes the lower order terms. We
will not address the lower order terms in $e$ further. For the following discussion, we assume a two dimensional tree
for concreteness. Higher dimensional trees will be addressed in the final section. We also assume a flop rate $r$ for
the machine. In order to calculate the coefficients in~\eqref{eq:complexity}, we will analyze the test problem from
vortex fluid dynamics shown in Section~\ref{sec:problem}.

    The first term in~\eqref{eq:complexity} subsumes all perfectly parallel work in the FMM, namely the initialization
of multipole expansions, the evaluation of local expansions on the finest tree level, and the final sum of multipole and
direct contributions. From above, our test problem uses the Biot-Savart kernel in two dimensions, and complex numbers to
represent positions, and we recall~\eqref{eq:regB-S2DGau}
\begin{equation*}
  \uKern_{\sigma}(z) = \frac{1}{2 \pi |z|^{2}}(-v, u) \left(1-\exp\left(-\frac{|z|^{2}}{2\sigma^{2}}\right)\right).
\end{equation*}
We carry out a $t$ term multipole expansion in which the $m$th coefficient is given by
\begin{equation}
  \frac{-i}{2\pi} \sum_{\mathrm{particles}} \Gamma_p \left( \vec{x}_p - \vec{x}_c \right)^m
\end{equation}
where $\Gamma_p$ is the circulation of particle $p$ and $\vec{x}_p$ its position, and $\vec{x}_c$ the cell center. For
details, please refer to~\cite{CruzBarbaKnepley2009}. We may now calculate the constant $a$ precisely. We will split
the different contributions into pieces so that $\sum_i a_i = a$.


The work to initialize all multipole expansions to order $t$, assuming 6 flops per complex multiplication, is given by
\begin{eqnarray*}
  W_\mathrm{init} &=& 2 + \sum_b \sum_p \left(3 + (6 + 2)t\right) \\
                  &=& 2 + (8 t + 3) N
\end{eqnarray*}
where $b$ runs over boxes on the finest level and $p$ runs over particles in a given box $b$. We move the constant term
to the low order part $e$. Thus our total time for initialization is
\begin{equation}
  T_\mathrm{init} = \frac{(8 t + 3)}{r} \frac{N}{P}
\end{equation}
and we have
\begin{equation}
  a_1 = \frac{(8 t + 3)}{r}.
\end{equation}
Since, we calculate the local expansions coefficients using the tree structure, we need only produce the local series
$\left( \vec{x}_p - \vec{x}_c \right)^m$ in order to evaluate the contribution at each particle location. Thus we have
\begin{equation}
  a_2 = \frac{(8 t + 2)}{r}.
\end{equation}
and for the final summation
\begin{equation}
  a_3 = \frac{2}{r}.
\end{equation}
The perfectly parallel work is therefore
\begin{equation}
  T_1 = \frac{16 t + 7}{r} \frac{N}{P}.
\end{equation}

The second and third terms represent the development of multipole expansions for all tree cells by translation (M2M) and
addition of expansions in an upward pass through the tree. The work is constant per tree cell, so the total work is
given by
\begin{equation*}
  W_\mathrm{up} = \sum^{L-1}_{l=2} 4^l 4 (2 + 2t^2 + 4t(t-1)) = c_1 \sum^{L-1}_{l=2} 4^l.
\end{equation*}
However, when we look at the parallel work, after a certain level, $\log_4 P$, there are fewer tree cells than processes
and thus some processes become idle,
\begin{eqnarray}
  T_\mathrm{up} &=& c_1 \sum^{L-1}_{l=2} 4^l \\
                &=& c_1 \left( \sum^{L-1}_{l=\log_4 P} \frac{4^l}{P} + \sum^{\log_4 P}_{l=2} 1 \right) \\
                &=& c_1 \left( \frac{1}{P} \frac{4^L - 4^{\log_4 P}}{4 - 1} + \log_4 P - 2 \right) \\
                &=& c_1 \left( \frac{N}{3 B P} + \log_4 P - \frac{7}{3} \right).
\end{eqnarray}
We move the constant into the low order terms $e$, and will add the first term into the expression for $c$. The second
term represents the \emph{reduction bottleneck} and could limit the scalability of FMM. In fact, this is stated
by~\cite{GreengardGropp1990,GumerovDuraiswami2008} and all works known to the author. However, this conclusion ignores
potential concurrency in the algorithm which will be addressed in Section~\ref{sec:concurrency}.

    In order to quantify this potential concurrency, we will have to consider the fourth term, which describes the direct
interaction between particles in the same and neighboring tree cells. We will compute each particle pair twice, once
from each end, to eliminate memory contention issues at the cost of additional flops. The work done depends on the
number of neighboring boxes, so we have three terms
\begin{eqnarray*}
  W_\mathrm{direct} &=& 22 \left( 4 (4 B^2 - B) + (2^{L+2} - 8) (6 B^2 - B) + (4^L - 2^{L+2} + 4) (9 B^2 - B) \right) \\
                    &=& 22 \left(9 4^L B^2 - 3 (2^{L+2} - 8) B^2 - 20 B^2 - 4^L B \right) \\
                    &=& 22 \left(9 \frac{N}{B} B^2 - 12 \sqrt{\frac{N}{B}} B^2 + 4 B^2 - N \right)
\end{eqnarray*}
where, in the first line, the first term counts flops for the four corner boxes, the second the boxes along the four sides, and the third
count flops in the interior boxes. Here we use 9 flops for complex division and 1 flop for the exponential. The second
and third terms of the last line can be moved to the lower order $e$, while the fourth term can be used to correct the
perfectly parallel term
\begin{equation}
  a_4 = -\frac{22}{r}.
\end{equation}
The first term then gives the dominant contribution to the time
\begin{equation}
  T_4 = \frac{198}{r} \frac{NB}{P},
\end{equation}
so that $d = 198/r$.



The third term involves three parts: a transformation of the multipole expansion to a local expansion (M2L) in each cell, a
reduction of the new local expansions for each cell, and a translation of the full local expansion to child cells. The
M2L transformation and reduction does work
\begin{eqnarray*}
  W_\mathrm{M2L} &=& \sum^{L}_{l=2} 4^l 27 (2 + 2 t^2 + 15 t^2) \\
                 &=& 27 (2 + 2 t^2 + 15 t^2) \frac{4^{L+1} - 4^2}{4 - 1} \\
                 &=& 9 (2 + 2 t^2 + 15 t^2) \left( 4^{L+1} - 16 \right)
\end{eqnarray*}
where we have used 27 as the maximum interaction list size. Since the work done by cells with smaller interaction lists
will be smaller by a factor of $N^{-1/2}$, we neglect them here. The L2L translations do work
\begin{eqnarray*}
  W_{L2L} &=& \sum^{L}_{l=3} 4^l 4 (2 + 2 t^2 + 8 t^2) \\
          &=& \frac{4}{3} (2 + 2 t^2 + 8 t^2) \left( 4^{L+1} - 64 \right).
\end{eqnarray*}
We can now give the time estimate
\begin{equation}
  T_\mathrm{down} = \frac{4}{3} \frac{26 + 26 t^2 + 167 t^2}{r} \frac{N}{B P}
\end{equation}
where we have discarded lower order terms, so that we have
\begin{equation}
  c_2 = \frac{104 + 772 t^2}{3 r}.
\end{equation}

\begin{table}
\begin{center}
\begin{tabular}{|l|c|}
\hline
a & $\frac{1}{r} \left( 16 t - 15 \right)$ \\
\hline
b & $\frac{1}{3 r} \left( 84 - 48 t + 571 t^2 \right)$ \\
\hline
c & $\frac{1}{3 r} \left( 128 - 48 t + 844 t^2 \right)$ \\
\hline
d & $\frac{1}{r} 198$ \\
\hline
\end{tabular}
\caption{Coefficients for the complexity expression in~\eqref{eq:complexity} in terms of the flop rate $r$ and the
expansion order $t$.}
\label{tab:coeff}
\end{center}
\end{table}

Thus, for the specific case of our test problem, we can give values for all coefficients in the complexity
estimate of~\eqref{eq:complexity} in terms of the flop rate $r$, as shown in Table~\ref{tab:coeff}.

\section{Concurrency}\label{sec:concurrency}


    Our hypothesis is that computation of the local direct interaction among neighboring particles can be done
concurrently with multipole calculations on coarse grid levels in order to maintain full utilization of the machine. In
order to determine whether local interaction calculations can be used to cover a loss of concurrency at coarser grid
levels, we must decide how many particles will be allocated to each box. Greengard and Gropp determine the optimal
number of particles per box by minimizing the total time. Finding the minimum of~\eqref{eq:complexity}, they obtain
\begin{equation}
  B_{opt} = \sqrt{\frac{c}{d}} \approx 30,
\end{equation}
where they have used, in 2D, $c = 25 s^2$, $d = 9$, and in single precision $s = 15$.

    Following this analysis, Gumerov and Duraiswami~\cite{GumerovDuraiswami2008} determine $B_{opt}$ in the same manner, but with
slightly different constants so that
\begin{equation}
  B_{opt} \approx 91.
\end{equation}
However, they use 320 particles in each box for the examples in their paper. This will results in higher computational
performance, but suboptimal total running time. For our test problem, we have
\begin{equation}
  B_{opt} = \sqrt{\frac{128 - 48 t + 844 t^2}{594}} \approx 18
\end{equation}
for $t = 15$. If this exceeds the minimum $B$ necessary to cover the reduction bottleneck, we make no tradeoff in
reorganizing our computation for enhanced concurrency.


    Using our model from Section~\ref{sec:complexity}, we can balance the time in direct evaluation with idle time for
small grids. Assume that only a single thread, or thread group, works on the first $L_{root}$ tree levels, so that
$L_{root}$ is the level where we start to lose concurrency and can no longer occupy each thread with a separate box. The
total time for direct evaluation is given by
\begin{equation}
  d \frac{N B}{P},
\end{equation}
and the work on the root tree by
\begin{equation}
  b L_{root}.
\end{equation}
Thus, we need
\begin{equation}
  B \ge \frac{b}{d} \frac{P L_{root}}{N}
\end{equation}
in order to cover the bottleneck completely with direct evaluation calculations. Consulting Table~\ref{tab:coeff}, we
have
\begin{eqnarray}
  B &\ge& \frac{84 - 48 t + 571 t^2}{594} \frac{P L_{root}}{N} \\
    &\ge& 215.22 \frac{P L_{root}}{N} \\
\end{eqnarray}
where we have used $t = 15$. It is common to take $L_{root} = \log_4 P$ as in the GG model, so that
\begin{equation}
  B \ge 215.22 \frac{P \log_4 P}{N}.
\end{equation}
We see that in order to achieve complete overlap of these computations, memory must scale slightly faster than
linearly. Although this may ultimately limit scalability, even machines an order of magnitude larger than today will not
test this limit.

    If we take a typical problem with $N = 10^6$ and $P = 10^4$, we see that
\begin{equation}
  B \ge 14.30
\end{equation}
is enough to eliminate the barrier to scaling. Thus, the optimal $B$ is enough to guarantee optimal scaling of only a
million particles on a modern GPU cluster, such as the Tesla, if the computation is reorganized to allow this
overlap.

\begin{figure}\label{fig:fmmMinimumSize}
\begin{center}
  \includegraphics[width=4in]{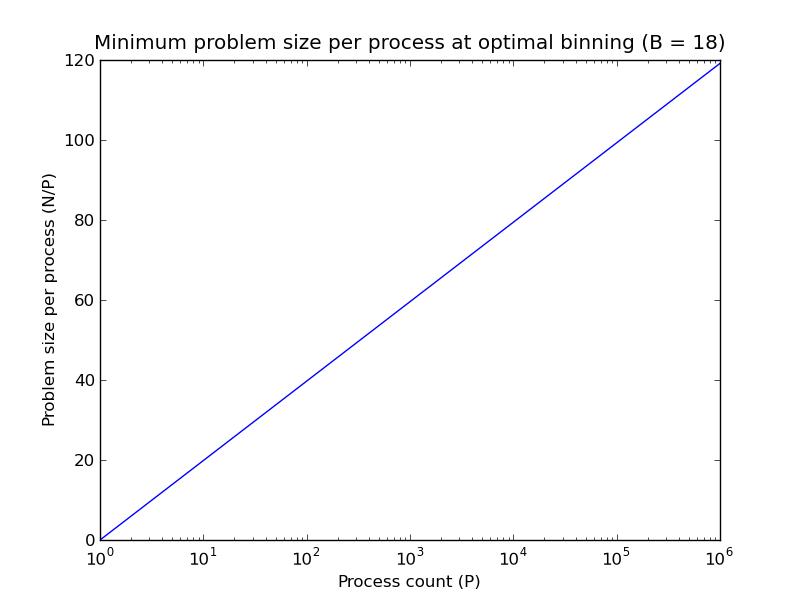}
\end{center}
\caption{The minimum problem size per process, $N/P$, for which no bottleneck will appear as a function of the number of
processes $P$.}
\end{figure}

    If we instead demand that $N/P$ is a fixed constant $M$, so that we have memory scalability, and fix $B$ at its
optimal value 18, we have
\begin{equation}
  M \ge 12 \log_4 P.
\end{equation}
Thus even a very large machine, with more than one million cores, would need to assign no more than 120 particles per
core. This can be seen clearly in Fig.~\ref{fig:fmmMinimumSize}, which shows the minimum problem size per process,
$N/P$, for which no bottleneck will appear as a function of the number of processes $P$.

\section{Conclusions}
\label{conclusions}

    We have shown that the potential bottleneck due to decreasing workload on higher levels of the FMM tree can be
alleviated by overlapping multipole computations with direct kernel evaluations on the finest level grid. The
implementation of this overlap on the NVIDIA Tesla architecture will be detailed in an upcoming publication.

    This analysis could be extended for more general scenarios, and we posit two immediately relevant examples. First,
uneven particle distributions will produce more direct computation for the same $N$ than the evenly distributed case, so
that the estimates remain valid and the bottleneck can be avoided. If instead the density of particles per box remains
constant, the estimates do not change and the bottleneck disappears as well. Second, in three or higher dimensions, the
balance of multipole work to direct computation will change. Since the size of the interaction list will increase
quickly, it is likely that minimum problem size to eliminate the bottleneck will increase, however the optimal number of
particle per box will also increase. Thus, a more detailed analysis will be necessary in this case.

\section{Vita}

Matthew G. Knepley received his B.S. in Physics from Case Western Reserve University in 1994, an M.S. in Computer
Science from the University of Minnesota in 1996, and a Ph.D. in Computer Science from Purdue University in 2000. He
was a Research Scientist at Akamai Technologies in 2000 and 2001. Afterwards, he joined the Mathematics and Computer
Science department at Argonne National Laboratory (ANL), where he was an Assistant Computational Mathematician, and a
Fellow in the Computation Institute at University of Chicago. In 2009, he joined the Computation Institute as a Senior
Research Associate. His research focuses on scientific computation, including fast methods, parallel computing, software
development, numerical analysis, and multicore architectures. He is an author of the widely used PETSc library for
scientific computing from ANL, and is a principal designer of the PetFMM and PetRBF libraries, for the parallel fast
multipole method and parallel radial basis function interpolation. He was a J.~T. Oden Faculty Research Fellow at the
Insitute for Computation Engineering and Sciences, UT Austin, in 2008.

\section{Acknowledgements}

This work was supported by the U.S. Dept. of Energy under Contract DE-AC01-06CH11357.

\bibliographystyle{elsarticle-num}
\bibliography{writings}

\begin{thebibliography}{1}
\expandafter\ifx\csname url\endcsname\relax
  \def\url#1{\texttt{#1}}\fi
\expandafter\ifx\csname urlprefix\endcsname\relax\def\urlprefix{URL }\fi
\expandafter\ifx\csname href\endcsname\relax
  \def\href#1#2{#2} \def\path#1{#1}\fi

\bibitem{GreengardGropp1990}
L.~Greengard, W.~D. Gropp, A parallel version of the fast multipole method,
  Comp.\ Math.\ Appl. 20~(7) (1990) 63--71.

\bibitem{GreengardRokhlin1987}
L.~Greengard, V.~Rokhlin, A fast algorithm for particle simulations, J.
  Comput.\ Phys. 73~(2) (1987) 325--348.
\newblock \href {http://dx.doi.org/10.1016/0021-9991}
  {\path{doi:10.1016/0021-9991}}.

\bibitem{Brandt77}
A.~Brandt, Multi-level adaptive solutions to boundary-value problems,
  Mathematics of Computation 31~(138) (1977) 333--390.
\newblock \href {http://dx.doi.org/10.2307/2006422}
  {\path{doi:10.2307/2006422}}.

\bibitem{StroutCarterFerranteFreemanKreaseck02}
M.~M. Strout, L.~Carter, J.~Ferrante, J.~Freeman, B.~Kreaseck, Combining
  performance aspects of irregular gauss-seidel via sparse tiling, in: 15th
  Workshop on Languages and Compilers for Parallel Computing (LCPC), LNCS,
  College Park, Maryland, 2002, pp. 90--110.
\newblock \href {http://dx.doi.org/10.1007/11596110_7}
  {\path{doi:10.1007/11596110_7}}.

\bibitem{Pozrikidis07}
C.~Pozrikidis, Fluid Dynamics: Theory, Computation, and Numerical Simulation,
  Kluwer (Springer), 2001.

\bibitem{CruzBarba2009}
F.~A. Cruz, L.~A. Barba, Characterization of the accuracy of the fast multipole
  method in particle simulations, Int.\ J. Num.\ Meth.\ Eng. 79~(13) (2009)
  1577--1604.
\newblock \href {http://dx.doi.org/10.1002/nme.2611}
  {\path{doi:10.1002/nme.2611}}.

\bibitem{CruzBarbaKnepley2009}
F.~A. Cruz, L.~A. Barba, M.~G. Knepley, {PetFMM}\,---\,a dynamically
  load-balancing parallel fast multipole library, submitted; preprint on
  \href{http://arxiv.org/abs/0905.2637}{http://arxiv.org/abs/0905.2637} (2009).

\bibitem{GumerovDuraiswami2008}
N.~A. Gumerov, R.~Duraiswami, Fast multipole methods on graphics processors, J.
  Comp.\ Phys. 227~(18) (2008) 8290--8313.
\newblock \href {http://dx.doi.org/doi:10.1016/j.jcp.2008.05.023}
  {\path{doi:doi:10.1016/j.jcp.2008.05.023}}.

\end{thebibliography}

\end{document}